\documentclass[aps,prl,twocolumn,showpacs,superscriptaddress,groupedaddress,nofootinbib]{revtex4}
\usepackage{xcolor}
\usepackage{graphicx,color,bm}

\usepackage{amsmath,amssymb}
\usepackage{epstopdf}
\usepackage{ulem}
\usepackage{array}
\usepackage{verbatim}
\usepackage[utf8]{inputenc}
\usepackage{rotating}
\newcolumntype{K}[1]{>{\centering\arraybackslash}m{#1}}

\usepackage[colorlinks,linkcolor=blue,citecolor=blue]{hyperref}

\newcommand{\orcid}[1]{\href{https://orcid.org/#1}{\,\includegraphics[width=8px]{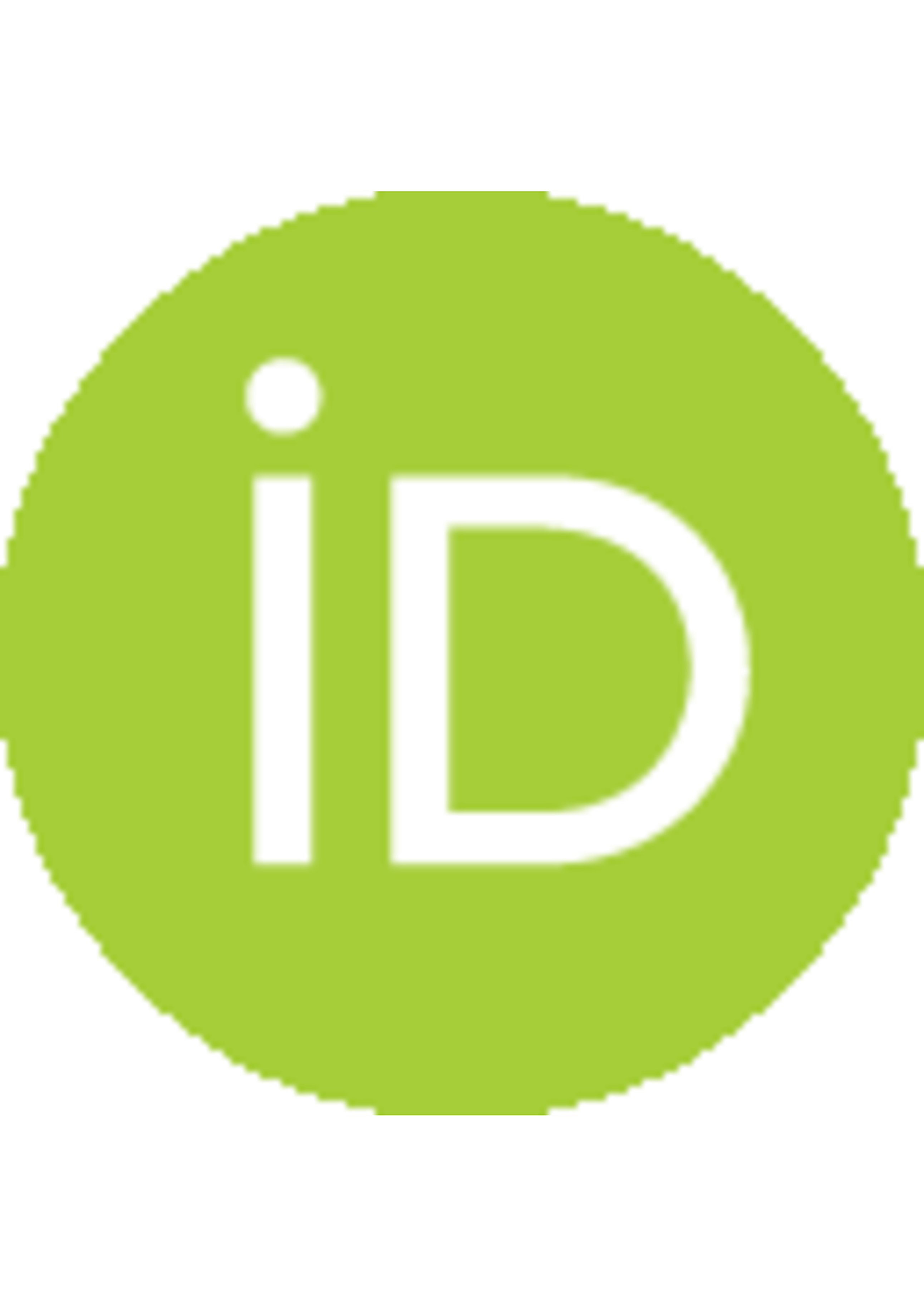}}}

\usepackage{mathtools}

\usepackage{tabularx}

\begin{document}

\title{
No evidence for phantom crossing: local goodness-of-fit improvements\\ do not persist under global Bayesian model comparison
}

\author{Bikash R. Dinda \orcid{0000-0001-5432-667X}}
\email{bikashrdinda@gmail.com}
\affiliation{Department of Physics \& Astronomy, University of the Western Cape, Cape Town 7535, South Africa}

\author{Roy Maartens \orcid{0000-0001-9050-5894}}
\email{rmaartens@uwc.ac.za}
\affiliation{Department of Physics \& Astronomy, University of the Western Cape, Cape Town 7535, South Africa}
\affiliation{National Institute for Theoretical \& Computational Science, Cape Town 7535, South Africa}

\author{Shun Saito \orcid{0000-0002-6186-5476}}
\email{saitos@mst.edu}
\affiliation{Institute for Multi-messenger Astrophysics \& Cosmology, Department of Physics, Missouri University of Science \& Technology,  
Rolla, MO 65409, United States of America}
\affiliation{Kavli Institute for the Physics \& Mathematics of the Universe, University of Tokyo, Kashiwa, Chiba 227-8583, Japan}


\begin{abstract}
Recent cosmological data have been interpreted as indicating deviations from $\Lambda$CDM within the standard $w_0w_a$ parametrization, including hints of phantom crossing and dynamical dark energy. However, such inferences can be parametrization-dependent and need not imply a statistically robust detection. We test these claims by comparing $\Lambda$CDM, $w_0w_a$, and thawing quintessence models, using the Deviance Information Criterion (DIC) and the Bayesian evidence $\ln \mathcal{Z}$. We find that $w_0w_a$ can provide a slightly improved local fit; however, this improvement is confined to a limited region of the parameter space. The global Bayesian evidence does not support it once the full prior volume is accounted for. In particular, cases with $\Delta{\rm DIC}<0$ but $\Delta \ln \mathcal{Z}<0$ indicate that these improvements are not statistically significant. We show that all models are statistically indistinguishable, and that there is no statistically consistent evidence across different datasets for either dynamical dark energy or phantom crossing.
\end{abstract}


\maketitle
\date{\today}

The baryon acoustic oscillation (BAO) data from the Dark Energy Spectroscopic Instrument (DESI) (data release 1, DR1 \citep{DESI:2024mwx,DESI:2024aqx,DESI:2024kob} and  DR2 \citep{DESI:2025zgx,DESI:2025fii}), combined with Planck 2018 cosmic microwave background (CMB) data \citep{Planck:2018vyg,ACT:2023kun}, and type Ia supernova (SNIa) data \citep{Brout:2022vxf,Rubin:2023jdq,DES:2024jxu}, have suggested evidence for dynamical dark energy. For example, CMB and DESI DR2 combined with Pantheon+ \citep{Brout:2022vxf}, Union3 \citep{Rubin:2023jdq}, Dark Energy Survey Year 5 (DES Y5) \citep{DES:2024jxu} and DES Dovekie \citep{DES:2025sig}, have $3.8\sigma$, $2.8\sigma$, $4.2\sigma$ and $3.2\sigma$ deviations respectively from the $\Lambda$CDM model. Here $\sigma$ denotes the Gaussian-equivalent frequentist significance of the deviation of the best-fit $(w_0,w_a)$ from the $\Lambda$CDM point $(w_0,w_a)=(-1,0)$. These are computed using the Mahalanobis distance in the $(w_0,w_a)$ plane with the full covariance matrix. The standard $w_0w_a$ parametrization of the equation of state of dark energy is
\begin{equation}
w(z)=w_0+w_a\, \frac{z}{1+z} \, ,
\label{eq:1}
\end{equation}
where $z$ is the redshift.

\begin{figure}[!htbp]
\centering
\includegraphics[width=1.1\linewidth]{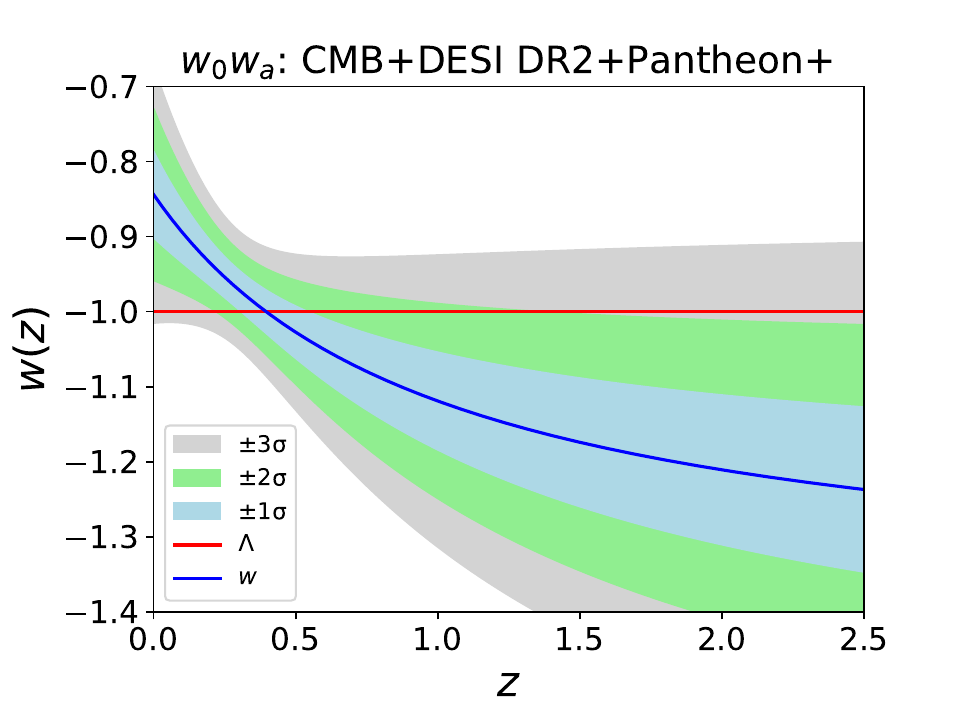}
\caption{
$w(z)$ from CMB+DESI DR2+Pantheon+ in the standard $w_0w_a$ parametrization, \autoref{eq:1}.
}
\label{fig:1}
\end{figure}

Specifically, the deviation from the standard $\Lambda$ model, $w(z)=-1$, has the feature that at early times ($z>0.4$), dark energy is phantom ($w<-1$) and at late times ($z<0.4$) it is non-phantom ($w>-1$), leading to a phantom crossing at $z\sim\! 0.4$ \citep{Shlivko:2026jxa,Keeley:2025rlg,Gomez-Valent:2025mfl,Guedezounme:2025wav}. Also, see \citep{Antusch:2026ldp,Roy:2025cxk,Silva:2025twg,Gialamas:2025pwv,Avila:2025phf,Wang:2025vfb,Wang:2026fuh,Jiang:2024xnu,Wang:2025znm,Wang:2026wrk,Li:2025ops} for related discussions.
\autoref{fig:1} illustrates this with the 
best-fit values obtained from the CMB+DESI DR2+Pantheon+ combination of data,
$w_0=-0.844 \pm 0.057$, $w_a=-0.55 \pm 0.22$ and normalized covariance between them, $\rho_{\rm Cov}(w_0,w_a)=-0.89$. For DESI DR2, we use Table~IV in \citep{DESI:2025zgx}. We use CMB information from the Planck 2018 release, including TT, TE, EE, low-$\ell$ polarization (lowE), and lensing measurements \citep{Planck:2018vyg}, in combination with Atacama Cosmology Telescope (ACT) DR6 CMB lensing data \citep{ACT:2023kun,ACT:2023dou}, corresponding to the baseline likelihood combination \texttt{base-plikHM-TTTEEE-lowl-lowE-lensing}. Instead of employing the full CMB likelihood, we adopt a compressed likelihood (CMB distance priors), described by three parameters: the physical baryon density $\Omega_{b0}h^2$, the physical matter density $\Omega_{m0}h^2$, and the angular size of the sound horizon at recombination $\theta_* = r_*/D_M(z_*)$, where $r_*$ is the comoving sound horizon at photon decoupling $z_*$, $D_M(z)$ is the comoving angular diameter distance, and the redshift of last scattering is $z_* \simeq 1089$. The mean values and covariance matrix of these parameters are taken from the main paper of the DESI DR2 dark energy analysis by \citep{DESI:2025zgx} (see their Appendix A). Additionally, assuming standard early-Universe physics (fixed effective number of relativistic species and CMB temperature), we fix the radiation density to $\Omega_{r0}h^2 = 4.15327 \times 10^{-5}$ \citep{Chen:2018dbv,Zhai:2019nad,Dinda:2024kjf}. We call this combination `CMB' here.
The approximation is validated in Appendix A and Figure 14 of \citep{DESI:2025zgx}, which showed a negligible difference in the central values of $(w_{0},w_{a})$ and $\sim\! 0.7\sigma$ difference in the contour size due to the lensing contribution in the full CMB likelihood.
Another point to notice from \autoref{fig:1} is that the non-phantom region is excluded at $\sim\! 2 \sigma$ from CMB+DESI DR2+Pantheon+.

\begin{figure}[!htbp]
\centering
\includegraphics[width=1.05\linewidth]{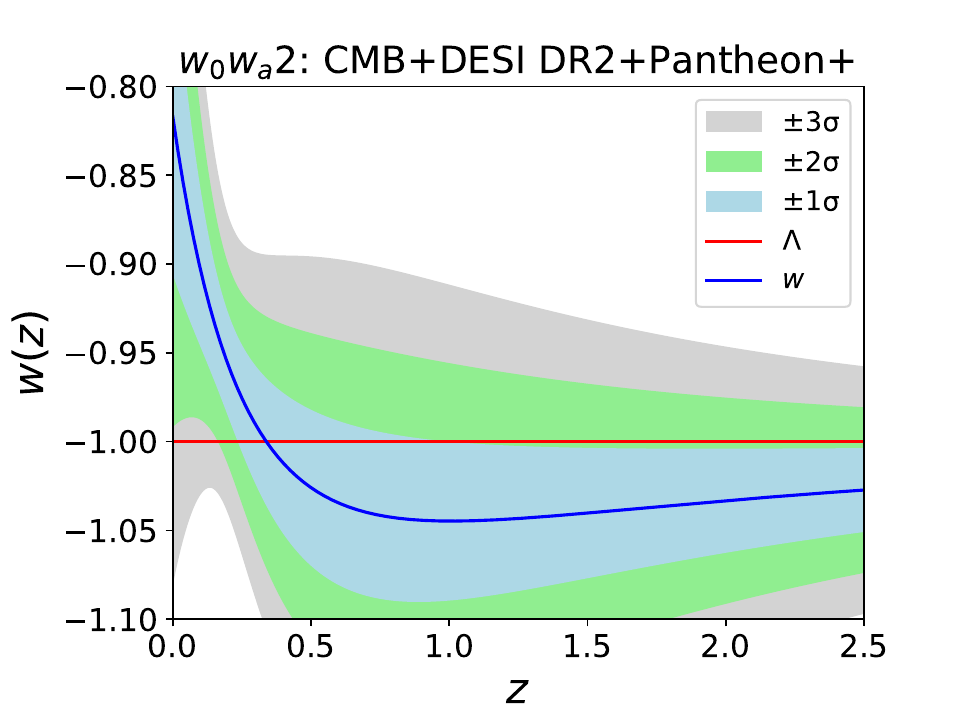}
\caption{
$w(z)$ from CMB+DESI DR2+Pantheon+ in the alternative parametrization, \autoref{eq:2}.
}
\label{fig:2}
\end{figure}

However, the standard $w_0w_a$ or similar parametrizations may not provide a reliable inference of the deviation from  $\Lambda$CDM \citep{Dinda:2025svh,Dinda:2024kjf} -- because the viable regions in the $(w_0,w_a)$ plane are not unique and can differ across dark energy parametrizations \citep{Akarsu:2026pom,Montefalcone:2026iga,Nesseris:2025lke,Xu:2026sbw}. This can lead to parametrization-dependent and potentially biased conclusions. To demonstrate this fact, we consider an alternative parametrization, $w_0w_a 2$: 
\begin{equation}
w(z)=-1+\frac{1+w_0+(3+3w_0+w_a)z}{(1+z)^3} \, .
\label{eq:2}
\end{equation}
In this alternative parametrization, the parameter constraints  are $w_0=-0.817 \pm 0.087$, $w_a=-1.09 \pm 0.66$, and $\rho_{\rm Cov}(w_0,w_a)=-0.94$ from CMB+DESI2+Pantheon+. The corresponding $w(z)$ evolution is displayed in \autoref{fig:2}, which shows that the non-phantom region is excluded only at $\sim\! 1\sigma$ -- compared to $\sim\! 2\sigma$ for the standard $w_0w_a$.

\begin{figure}[!htbp]
\centering
\includegraphics[width=1.0\linewidth]{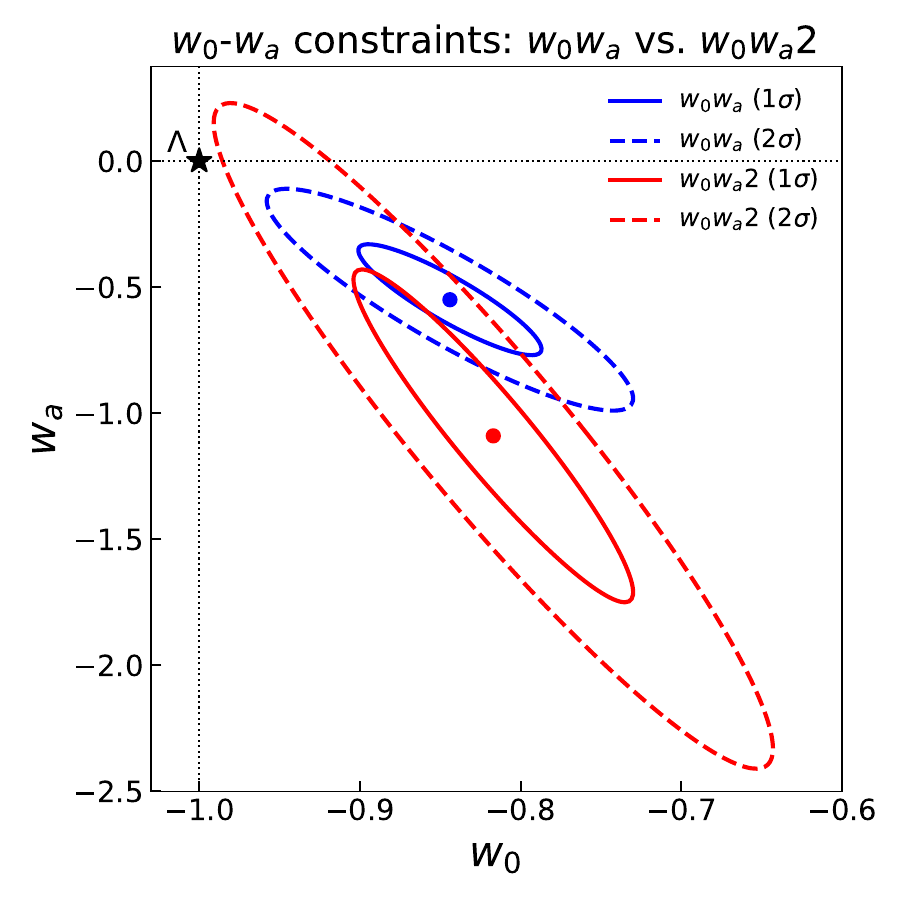}
\caption{
Confidence contours for the $w_0w_{a}$ and $w_0w_{a}2$ parametrizations using CMB+DESI DR2+Pantheon+ data.
}
\label{fig:3}
\end{figure}

Next, we compare the $1\sigma$ and $2\sigma$ contour ellipses in the $(w_0,w_a)$ plane in \autoref{fig:3}. Clearly, the contour regions are different. In the alternative $w_0w_a2$ parametrization, $\Lambda$CDM is 2.3$\sigma$ away, compared to 2.8$\sigma$ for the standard case. An apparent deviation from $\Lambda$CDM in a given parametrization does not necessarily imply a statistically robust detection of dynamical dark energy.

To establish robust evidence of dynamical dark energy and phantom crossing, it is essential to assess whether the parametrization itself is preferred over  $\Lambda$CDM using quantitative model selection criteria, such as the Deviance Information Criterion (DIC) and the Log Bayesian evidence $\ln \mathcal{Z}$ \citep{Ong:2026tta,Hergt:2026moc}. DIC combines a measure of goodness-of-fit with a certain penalty for model complexity, where the effective number of parameters is estimated from the posterior distribution. In this work, DIC is computed from the MCMC posterior obtained with \texttt{emcee} \citep{Foreman-Mackey:2012any}, with averages taken over the chains and the reference point set to the posterior mean. In contrast,  $\ln \mathcal{Z}$ is computed using nested sampling via \texttt{PyMultiNest} \citep{Feroz:2008xx}, which naturally incorporates an Occam penalty by accounting for the full prior volume. We interpret $|\Delta \ln \mathcal{Z}|$ using the Jeffreys scale: values below 1 are inconclusive, 1--2.5 indicate weak evidence, 2.5--5 moderate evidence, and values greater than 5 indicate strong evidence. For DIC, we adopt the conventional scale: $|\Delta{\rm DIC}| < 2$ (inconclusive), $2$--$5$ (weak), $5$--$10$ (moderate), and $>10$ (strong preference). Thus, while DIC evaluates goodness-of-fit with an effective penalty based on the posterior, $\ln \mathcal{Z}$ provides a global model comparison sensitive to the prior volume, which can lead to mild differences in preference when the improvement in fit is not significant. Model 1 is preferred over model 2 if $\Delta$DIC = DIC (model 1) - DIC (model 2) is negative, whereas $\Delta \ln \mathcal{Z}$ = $\ln \mathcal{Z}$ (model 1) - $\ln \mathcal{Z}$ (model 2) is positive.

\begin{table}
\centering
\caption{Comparison of the  $w_0w_a$ parametrization against  $\Lambda$CDM  through  $\Delta$DIC and $\Delta \ln \mathcal{Z}$.
}
\label{table:1}
\small
\begin{tabular}{l c c}
\hline\hline
Data & $\Delta$DIC & $\Delta \ln \mathcal{Z}$ \\
\hline
CMB+DESI DR2 & $-10.9$ & $-0.9$ \\
CMB+Pantheon+ & $+1.5$ & $-5.0$ \\
CMB+DESI DR2+Pantheon+ & $-3.8$ & $-2.9$ \\
CMB+DESI DR2+Union3 & $-10.7$ & $+0.8$ \\
CMB+DESI DR2+DES Y5 & $-26.5$ & $+8.1$ \\
CMB+DESI DR2+DES Dovekie & $-7.7$ & $-1.4$ \\
\hline\hline
\end{tabular}
\end{table}

The results for these two model selection criteria are shown in \autoref{table:1}. Interestingly, we find contradictory results from $\Delta$DIC and $\Delta \ln \mathcal{Z}$ in most cases  (see also the next tables). But these contradictions are not for strong preferences, but only for weak to moderate preferences. For example, for CMB+DESI DR2+DES Y5,  both criteria agree that the  $w_0w_a$ parametrization is strongly favoured over  $\Lambda$CDM and consequently is evidence for dynamical dark energy. For CMB+Pantheon+, the $\Lambda$CDM model is marginally strongly favoured over $w_0w_a$, analyzed through $\Delta \ln\mathcal{Z}$. For other cases, except for CMB+DESI DR2+Union3, although $\Delta$DIC and $\Delta \ln\mathcal{Z}$ provide opposite evidence, they are not strong enough to prefer one model over another. Especially, the cases where $\Delta{\rm DIC}<0$ but $\Delta \ln \mathcal{Z}<0$ indicate that, although the $w_0w_a$ parametrization provides a slightly better fit to the data, the improvement is confined to a limited region of parameter space (locally near the posterior peak) and does not compensate for the larger prior volume (globally). Consequently, the Bayesian evidence favours the $\Lambda$CDM model, and no robust preference for dynamical dark energy can be established. Hence, the $w_0w_a$ parametrization is not favoured over $\Lambda$CDM, and consequently, there is no evidence of dynamical dark energy for CMB+DESI DR2, CMB+DESI DR2+Pantheon+, and CMB+DESI DR2+DES Dovekie.

The results for CMB+DESI DR2+Union3 are interesting, because $w_0w_a$ is marginally strongly favoured over  $\Lambda$CDM  through DIC, but almost equivalent through $\ln\mathcal{Z}$. This indicates that the $w_0w_a$ parametrization provides an improved fit relative to $\Lambda$CDM in a limited region of the $(w_0,w_a)$ plane. Nevertheless, this improvement is insufficient to yield a statistically significant preference when the full parameter volume is considered. This further means that the $w_0w_a$ parametrization is insufficient to draw any conclusions about dynamical dark energy from the CMB+DESI DR2+Union3 data combination. Consequently, this opens a door for studying other parametrizations or theoretical models.

The results of comparison between $w_0w_a$CDM and $\Lambda$CDM are consistent with \citep{Ong:2026tta}. Also, for CMB+DESI DR2+DES Dovekie, our results are consistent with those of \citep{DES:2025sig}. Furthermore, the qualitative discussions are consistent with \citep{Xu:2026sbw,Lee:2025ysg}.

Next, we consider thawing quintessence models. Quintessence models are arguably the best physically self-consistent models of dark energy \citep{Dinda:2025iaq,Sultana:2026ych,Ouardi:2026nlf}. We consider two potentials, 
\begin{eqnarray}
\text{ExpTQ:} \quad && V(\phi) = V_0 \exp\Big[{-\lambda \Big(\frac{\phi}{M_{\rm Pl}}\Big)}\Big] \, ,
\label{eq:3} \\
\text{SqrtTQ:} \quad && V(\phi) = V_0 \Big[{1+\alpha \Big( \frac{\phi-\phi_0}{M_{\rm Pl}} \Big) }\Big]^{1/2} \, ,
\label{eq:4}
\end{eqnarray}
where $M_{\rm Pl}=1/\sqrt{8 \pi G}$ is the reduced Planck mass.
Additionally, we consider the   parametrization 
\begin{equation}
\text{ParamTQ:} \quad w(z) = -1 + \frac{3 (1+w_0)}{3(1 + z + z^2) + z^3} \, ,
\label{eq:5}
\end{equation}
which mimics the dynamics of thawing quintessence for $w_0>-1$.

\begin{table}
\centering
\caption{Quintessence models: comparison against $\Lambda$CDM.}
\label{table:2}
\small
\begin{tabular}{l l r r}
\hline\hline
Data & Model & $\Delta$DIC & $\Delta \ln\mathcal{Z}$ \\
\hline

CMB+DESI DR2 
& ExpTQ   & $-9.8$ & $-0.1$ \\
& SqrtTQ  & $-7.5$  & $+0.1$ \\
& ParamTQ & $+1.7$ & $-1.7$ \\

CMB+Pantheon+ 
& ExpTQ   & $-5.6$ & $+0.1$ \\
& SqrtTQ  & $-8.5$  & $+0.2$ \\
& ParamTQ & $+1.4$ & $-1.9$ \\

CMB+DESI DR2+Pantheon+ 
& ExpTQ   & $-5.4$ & $+1.1$ \\
& SqrtTQ  & $-9.6$  & $+1.2$ \\
& ParamTQ & $-0.8$ & $-1.0$ \\

CMB+DESI DR2+Union3 
& ExpTQ   & $-10.2$ & $+1.7$ \\
& SqrtTQ  & $-8.8$  & $+2.0$ \\
& ParamTQ & $-1.9$ & $-0.1$ \\

CMB+DESI DR2+DES Y5 
& ExpTQ   & $-16.5$ & $+7.7$ \\
& SqrtTQ  & $-18.7$  & $+8.4$ \\
& ParamTQ & $-15.0$ & $+6.1$ \\

CMB+DESI DR2+DES Dovekie 
& ExpTQ   & $-8.5$ & $+1.8$ \\
& SqrtTQ  & $-6.5$  & $+2.0$ \\
& ParamTQ & $-2.7$ & $-0.2$ \\

\hline\hline
\end{tabular}
\end{table}

We compute DIC and $\ln \mathcal{Z}$ for these three models and compare with $\Lambda$CDM in \autoref{table:2}. It is apparent that there is no strong evidence for thawing quintessence models over $\Lambda$CDM, and consequently, no evidence for dynamical dark energy, except for the CMB+DESI DR2+DES Y5 data combination. This combination has strong preferences for all three thawing quintessence models over  $\Lambda$CDM and hence strong evidence for dynamical dark energy within this dataset alone, as in the case of the $w_0w_a$ parametrization. This data combination is therefore an outlier in our analysis: both $w_0w_a$ and thawing quintessence are strongly favored over $\Lambda$CDM. However, the comparison between $w_0w_a$ and quintessence models (\autoref{table:3}) shows no statistically significant preference between phantom and non-phantom behavior. Thus, even in this case, the data do not provide robust evidence for phantom crossing.

\begin{table}
\centering
\caption{Quintessence models: comparison against  $w_0w_a$.}
\label{table:3}
\small
\begin{tabular}{l l r r}
\hline\hline
Data & Model & $\Delta$DIC & $\Delta \ln\mathcal{Z}$ \\
\hline

CMB+DESI DR2
& ExpTQ   & $+1.1$ & $+0.8$ \\
& SqrtTQ  & $+3.4$  & $+1.0$ \\
& ParamTQ & $+12.6$ & $-0.8$ \\

CMB+Pantheon+ 
& ExpTQ   & $-7.1$ & $+5.1$ \\
& SqrtTQ  & $-10.0$  & $+5.2$ \\
& ParamTQ & $-0.1$ & $+3.1$ \\

CMB+DESI DR2+Pantheon+ 
& ExpTQ   & $-1.6$ & $+4.0$ \\
& SqrtTQ  & $-5.8$  & $+4.1$ \\
& ParamTQ & $+3.0$ & $+1.9$ \\

CMB+DESI DR2+Union3 
& ExpTQ   & $+0.5$ & $+0.9$ \\
& SqrtTQ  & $+1.9$  & $+1.2$ \\
& ParamTQ & $+8.8$ & $-0.9$ \\

CMB+DESI DR2+DES Y5 
& ExpTQ   & $+10.0$ & $-0.4$ \\
& SqrtTQ  & $+7.8$  & $+0.3$ \\
& ParamTQ & $+11.5$ & $-2.0$ \\

CMB+DESI DR2+DES Dovekie 
& ExpTQ   & $-0.8$ & $+3.2$ \\
& SqrtTQ  & $+1.2$  & $+3.4$ \\
& ParamTQ & $+5.0$ & $+1.2$ \\

\hline\hline
\end{tabular}
\end{table}

In light of this, it is necessary to compare the $w_0w_a$ parametrization directly with thawing quintessence models, to assess whether the data distinguishes between phantom and non-phantom behavior. This comparison is relevant not only for the CMB+DESI DR2+DES Y5 combination, but also for cases where dynamical dark energy is weakly to moderately preferred (e.g., CMB+DESI DR2+Union3). We therefore perform model comparison and report the results in \autoref{table:3}.

We see from \autoref{table:3} that for CMB+DESI DR2+DES Y5, the quintessence and $w_0w_a$ models have almost equal evidence, and consequently, no conclusion can be made for the evidence of phantom versus non-phantom. Thus, thawing quintessence models are equally viable. Interestingly, we see that for CMB+Pantheon+, the scalar field thawing quintessence models are marginally strongly preferred over the standard $w_0w_a$ parametrization. Also, we find that, for the CMB+DESI DR2+Pantheon+ and CMB+DESI DR2+DES Dovekie data combinations, scalar field thawing quintessence models ExpTQ and SqrtTQ are moderately preferred over the standard $w_0w_a$ parametrization. Although not strongly evident, this indicates a moderate preference for non-phantom over phantom dark energy.

A further point to highlight here is that the parametrized thawing quintessence is not good enough to distinguish between phantom and non-phantom evidence. This may indicate that it is better to study thawing quintessence models directly through the scalar field evolution, instead of via a mimicking parametrization, especially for the evidence of the dynamical dark energy. This point deserves further investigation.
\begin{figure}[!htbp]
\centering
\includegraphics[width=1.05\linewidth]{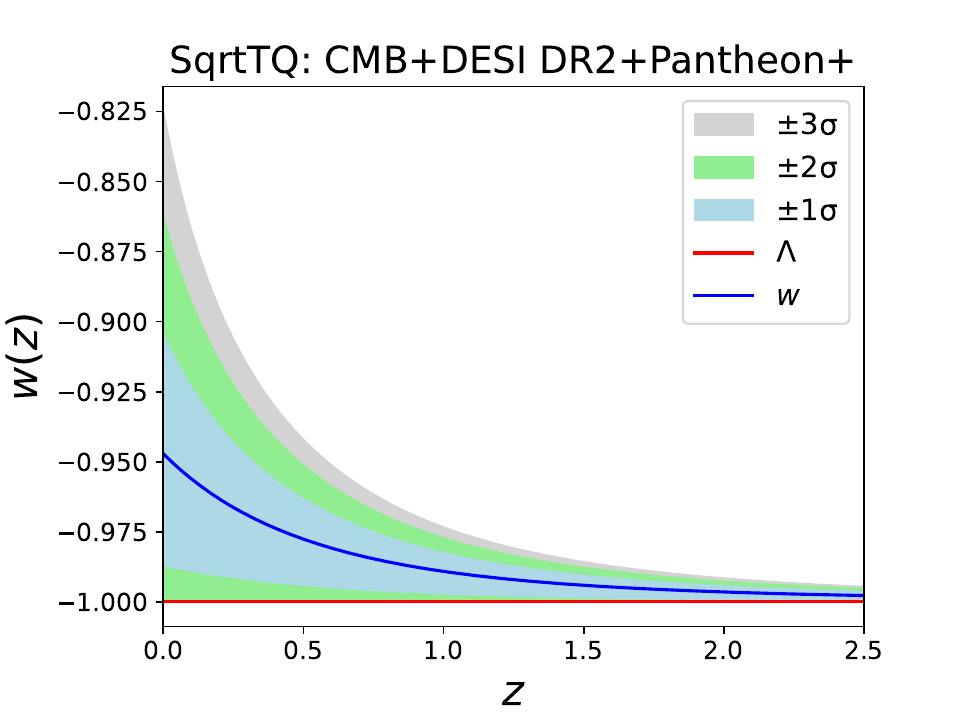}
\caption{
Best-fit $w(z)$ from CMB+DESI DR2+Pantheon+ in the scalar field thawing quintessence model with square-root potential, \autoref{eq:4}.
}
\label{fig:4}
\end{figure}

Next, we present in \autoref{fig:4} a similar plot to that in \autoref{fig:1} for the scalar field thawing quintessence model with a square-root potential, \autoref{eq:4}, for the CMB+DESI DR2+Pantheon+ data combination. Comparing the DIC and $\ln\mathcal{Z}$ in \autoref{table:3}, we see that the $w(z)$ behaviour in \autoref{fig:4} is moderately favoured over the behaviour in \autoref{fig:1}. This again demonstrates that no phantom crossing is evident, and that thawing quintessence models remain consistent with current data.

\begin{table}[t]
\centering
\caption{
Dependence of
$\Delta \ln \mathcal{Z}
=
\ln \mathcal{Z} (w_0w_a)
-
\ln \mathcal{Z} (\Lambda)$
on the adopted flat prior ranges on $w_0$ and $w_a$ for CMB+DESI DR2+Pantheon+.
}
\label{table:4}
\small
\begin{tabular}{c c c}
\hline\hline
$w_0$ prior & $w_a$ prior & $\Delta \ln \mathcal{Z}$ \\
\hline

$[-1.6,\,-0.4]$
& $[-3,\,3]$
& $-1.3$
\\

$[-2,\,0]$
& $[-3,\,3]$
& $-1.9$
\\

$[-2.5,\,0.5]$
& $[-5,\,5]$
& $-2.9$
\\

$[-3,\,1]$
& $[-5,\,5]$
& $-3.1$
\\

$[-3,\,1]$
& $[-10,\,10]$
& $-3.8$
\\

\hline\hline
\end{tabular}
\end{table}

\begin{figure}[!htbp]
\centering
\includegraphics[width=0.97\linewidth]{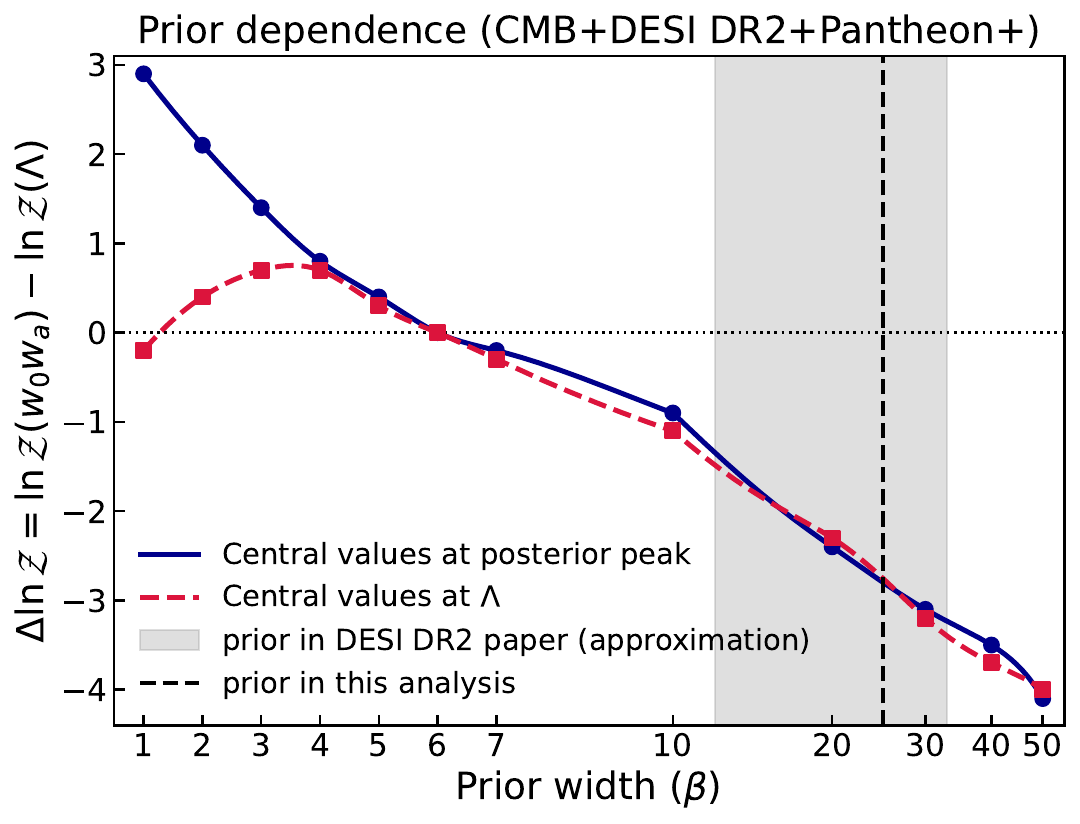}
\caption{
Dependence of $\Delta \ln \mathcal{Z}$ on prior widths of $w_0$ and $w_a$, parametrized by a single width parameter $\beta$, given in \autoref{pw}. Results are shown for priors centered on the posterior peak (solid blue) and on the $\Lambda$CDM point (dashed red).
}
\label{fig:5}
\end{figure}

Before presenting the conclusion, we must examine the dependence of the Bayesian evidence on the adopted priors. Since $\ln \mathcal{Z}$ is computed by integrating the likelihood over the full prior volume, it depends on the assumed prior ranges of the parameters \citep{Trotta:2008qt,Amendola:2024prl,Patel:2024odo,Gonzalez-Fuentes:2026rgu,Gonzalez-Fuentes:2025lei}. To test this dependence, we first vary the prior ranges of $w_0$ and $w_a$ for the CMB+DESI DR2+Pantheon+ data combination. The results are summarized in \autoref{table:4}. We find that $\Delta \ln \mathcal{Z}$ changes moderately with the choice of prior ranges. However, these changes do not modify the overall qualitative conclusion of the model comparison, provided the priors remain sufficiently broad. 

To understand this behavior in more detail, we also introduce a controlled scaling of the prior width using a single parameter $\beta$.  The corresponding results are shown in \autoref{fig:5}. We define the prior widths 
\begin{align}
& w_0^{\ast} - \beta \Delta w_0 \leq w_0 \leq w_0^{\ast} + \beta \Delta w_0 \; , \notag\\
& w_a^{\ast} - \beta \Delta w_a \leq w_a \leq w_a^{\ast} + \beta \Delta w_a \; ,
\label{pw}
\end{align}
where $(w_0^{\ast}, w_a^{\ast})$ denotes either the posterior maximum (solid blue) or the $\Lambda$CDM point (dashed red). We identify two distinct behaviors depending on the choice of the central point. If the priors are centered on the posterior peak, then decreasing $\beta$ (narrowing the prior) increases $\Delta \ln \mathcal{Z}$, since the high-likelihood region remains fully contained while the prior volume shrinks. In contrast, if the priors are centered on the $\Lambda$CDM point, the behavior is different. For moderate to high values of $\beta$, both the $\Lambda$CDM and posterior peak central values show similar trends. However, when $\beta$ becomes too small, the $\Lambda$CDM point moves outside the high-likelihood region. In this regime, $\Delta \ln \mathcal{Z}$ starts to decrease with decreasing $\beta$ and has a corresponding maximum. So, for intermediate values of $\beta$, the posterior is fully contained within the prior. In this regime, the evidence is stable and insensitive to moderate changes in prior width. This is the regime used for our main results. 

For very large $\beta$, the prior volume becomes very wide, and the evidence decreases due to the Occam penalty associated with unused parameter space. Overall, we should choose a region of priors that is neither too narrow nor extremely broad. Within this region, our conclusions do not depend on the precise choice of prior, provided it is physically reasonable and sufficiently broad to contain the posterior distribution.

Note that in \autoref{fig:5}, the curve based on central values at the expected posterior peak (blue) is included primarily for illustration, as it helps clarify the origin of the turnover and the appearance of a maximum in the case of $\Lambda$-centered priors. However, such a choice of central values is not meaningful in practice, since the posterior peak is not known a priori. The choice of central values is not significantly important, provided that the central values are not far from the $\Lambda$ point and the priors are sufficiently broad, since in that regime the results for different choices of central values overlap.

In \autoref{fig:5}, we highlight our main prior with the black dashed vertical line. We also include a grey region corresponding to the prior that is used in DESI DR2 \citep{DESI:2025fii}: since it can not be scaled with a single parameter like $\beta$, the actual vertical line is not possible to find, but it should be within the grey region.

Moreover, a general remark is that the dependence of the Bayesian evidence on priors becomes more prominently relevant when the preference from local goodness-of-fit is only at the level of about $3$--$4\sigma$. In this case, different model selection criteria, such as DIC and $\ln \mathcal{Z}$, may not always give consistent conclusions. This is seen in some of the data combinations considered here, while for stronger signals, such as those including DES Y5, the different criteria give consistent results.

Furthermore, for most of the data combinations, neither phantom ($w<-1$) nor non-phantom ($w>-1$) regions are strongly evident against each other or against $\Lambda$ ($w=-1$), and consequently, there is no strong evidence for dynamical dark energy. In general, model preference should be assessed primarily using the global (Log) Bayesian evidence $\ln \mathcal{Z}$, as it provides a robust comparison by accounting for the full parameter space and penalizing unnecessary model complexity. Taken together, we conclude that these data are still statistically consistent with physically motivated non-phantom models, such as thawing quintessence, and do not require parametrizations that permit phantom behavior.

\section*{Acknowledgements}
We thank Alan Heavens for his insightful comments and suggestions. We also thank Alex González-Fuentes for his useful comments and suggestions. BRD is supported by the South African Radio Astronomy Observatory and the National Research Foundation (Grant No. 75415). 
BRD, RM, and SS acknowledge support for this work from the University of Missouri South African Education Program (UMSAEP).
SS acknowledges support from the National Science Foundation under grants NSF-2219212 and NSF-2511145, and from NASA Grant \#80NSSC24M0021, ``Project Infrastructure for the Roman Galaxy Redshift Survey.'' 
SS is also supported in part by the World Premier International Research Center Initiative, MEXT, Japan.

\bibliographystyle{apsrev4-1}
\bibliography{references}

\end{document}